# High-Temperature Deformation Behavior of Co-Free Non-Equiatomic CrMnFeNi Alloy

F. J. Dominguez-Gutierrez[1], M. Frelek-Kozak[1], G. Markovic[2], M. A. Stróżyk[1], A. Daramola[3], M. Traversier[4], A. Fraczkiewicz[4], A. Zaborowska[1], T. Khvan[1], I. Jozwik[1], L. Kurpaska[1]

[1]*National Centre for Nuclear Research, NOMATEN CoE, ul. Andrzeja Soltana 7, 05-400 Świerk, Poland*
[2]*Institute for Technology of Nuclear and Other Mineral Raw Materials, 11000 Belgrade, Serbia*
[3]*University of Edinburgh, School of Physics and Astronomy, Edinburgh, Scotland*
[4]*Mines Saint-Etienne, Univ. Lyon, CNRS, UMR 5307 LGF, Saint-Etienne F-42023, France*

## Abstract

Cobalt-free high-entropy alloys (HEAs) have garnered interest for nuclear structural applications due to their good mechanical performance, thermal stability, and resistance to radiation-induced degradation, while avoiding long-lived Co radioisotopes. This study presents an experimental and computational investigation of the plastic deformation behavior of a non-equatomic CrMnFeNi alloy, designed to maintain a stability of fcc phase in a large domain of temperatures and to balance stacking fault (SF) energies for enhanced strain hardening and ductility. Tensile tests reveal a temperature-dependent reduction in mechanical strength, attributed to thermally activated deformation mechanisms and microstructural evolution. Molecular dynamics simulations of single- and polycrystals capture dislocation activity, SF formation, and twin nucleation as a function of strain and temperature. Electron backscatter diffraction (EBSD) confirms twin formation and grain boundary activity. The Schmid factor mapping is drawn to interpret local slip activity and anisotropic deformation behavior. The absence of Co leads to enhanced high-temperature strength compared to the Cantor alloy.

Corresponding author E-mail: javier.dominguez@ncbj.gov.pl

## 1. Introduction

The Cantor alloy (equiatomic CrMnFeCoNi) is a prototypical high-entropy alloy (HEA) that has garnered significant attention due to its good mechanical performance (attributed, especially at low temperatures, to activation of mechanical nanotwinning), thermal stability, and radiation resistance [1-7]. This face-centered cubic (fcc) alloy exhibits pronounced lattice distortion, sluggish diffusion kinetics, and robust solid-solution strengthening mechanisms [3-8]. These intrinsic attributes contribute to its superior strain hardening capacity, enhanced damage tolerance (especially at low temperatures), and excellent mechanical properties across a broad temperature spectrum, making it a prime candidate for demanding structural applications [4-16]. The Cantor alloy and many of other HEA developed on this basis contain Co, whose major role is to increase mechanical resistance and to stabilize the fcc phase [17]. Yet, Co becomes a strong gamma-emitter and forms long-lived radioisotopes under neutron irradiation [18-20], which makes it unsuitable for nuclear systems [21]. For this reason, intensive research efforts are being conducted searching for Co-free HEA alloys able to achieve a tailored set of properties. By decreasing the Co content often leads to a phase transformation from fcc to BCC, which may induce an unsuitable brittleness. HEAs, particularly those comprising transition metals such as Cr, Mn, Fe, and Ni, have emerged as promising candidates for structural applications due to their excellent mechanical strength, thermal stability, and resistance to harsh environments. However, the candidates HEA for nuclear applications need to avoid the formation of the σ phase (rich in Fe and Cr) which is highly undesirable because it significantly reduces the toughness, ductility, and corrosion resistance of the material, often leading to brittle fracture and premature failure in service [22,23]. To prevent σ phase formation, careful control of alloy composition is essential limiting high chromium and proper heat treatment strategies must be employed [22]. This includes avoiding prolonged exposure to critical temperature ranges, applying rapid cooling (quenching) after solution annealing, and incorporating stabilizing elements like nickel that inhibit σ phase precipitation. Among many works addressing this challenge, a $Cr_{15}Mn_{13}Fe_{39}Ni_{33}$ alloy was previously developed and made available for studies in the INNUMAT project [16]. This alloy has been shown to maintain fundamental similarities to the Cantor alloy and especially, its assets: low temperature twinning and phase stability (fcc) in large domain of temperatures [16-28], thanks to the presence of Mn and carefully controlled Ni content, avoiding excessive activation, while substituting Co's beneficial role via mechanisms like twinning- and transformation-induced plasticity, offering a cost-effective and adaptable solution for demanding environments [27-29].

In polycrystalline HEAs, the formation of twins during plastic deformation is a key mechanism for sustaining ductility while maintaining high strength. Twinning-assisted plasticity enhances work hardening capacity by introducing additional barriers to dislocation motion, delaying strain localization, and enabling more uniform deformation. The propensity for twinning is influenced by factors such as stacking fault energy, grain size, and strain rate, with finer grains typically promoting higher twin densities [30,31]. In high-temperature tensile testing, diffusion-assisted processes such as dislocation climb, grain boundary sliding, and dynamic recrystallization become increasingly relevant, impacting both ductility and strength [32]. While fcc HEAs generally exhibit stable mechanical performance over a wide temperature range, the onset of dynamic recovery and recrystallization at high temperatures can lead to strain softening, altering the balance between strength and ductility. Generally, the design of mechanical properties focuses on several deformation mechanisms, which depend on multiple factors, including crystallographic structure, temperature, stress fields, and others [1,2,33]. A clear or generalizable rule describing the influence of Co on mechanical properties has not been established. As Wu et al. [34] demonstrated, the presence of Co enhances performance at low temperatures. Furthermore, the authors reported higher work-hardening capabilities at cryogenic temperatures in Co-containing HEA systems, related to the deformation-induced nano-twinning effect [4]. However, it should be noted that similar properties can also be achieved in Co-free alloys [16, 17,35,36].

A comprehensive understanding of the deformation mechanisms governing HEAs behavior under elevated temperatures is essential for advancing materials' engineering applications. Electron Backscatter Diffraction (EBSD) is usually employed to characterize materials at post-deformation to assess grain structure evolution, crystallographic texture, and local strain distribution [37,38]. Computational modeling techniques such as molecular dynamics (MD) enable the exploration of atomic-scale interactions, dislocation dynamics, and defect evolution under mechanical loads [23,24,39-44]. These approaches complement each other by providing insights into the underlying mechanisms governing plastic deformation, such as dislocation slip, twinning, and grain boundary interactions [33,45]. Furthermore, the integration of simulations with experimental methodologies enhances predictive capabilities and accelerates the development of next-generation HEAs for high-performance applications [46,47]. Molecular dynamics (MD) offers time-resolved, atomistic insights into the deformation mechanisms that dictate the mechanical behavior of high-entropy alloys (HEAs), encompassing dislocation nucleation and glide, stacking-fault and twin formation, as well as stress- and temperature-induced phase transitions within chemically disordered, lattice-distorted matrices. Molecular dynamics correlates microscopic dynamics with macroscopic qualities such as elastic modulus, yield strength, ultimate tensile strength, ductility, and work hardening by analyzing the influence of strain rate, temperature, grain size, and composition on these phenomena. In this context, molecular dynamics (MD) improves experiments that may lack spatiotemporal resolution for nanoscale phenomena, enabling quantitative, mechanism-driven analysis of tensile properties in AlCoCrCuFeNi systems and AlxCoCrFeNi amorphous high-entropy alloys (HEAs), as well as the early stages of film growth (cluster nucleation/coarsening during annealing) [72-74].

In this work, we present an integrated experimental and computational investigation of the high-temperature plastic deformation behavior of a Co-free, non-equiatomic $Cr_{14.2}Mn_{12.5}Fe_{39.7}Ni_{33.6}$ alloy (HEA-1). While both the Cantor alloy and these novel Co-free compositions benefit from high configurational entropy and multi-element interactions, subtle differences in atomic bonding and stacking-fault energetics significantly affect their mechanical response, particularly under extreme conditions such as high radiation exposure or elevated temperatures. Thus, a family of Co-free Ni-Fe-Cr-Mn alloys was designed and synthesized experimentally at MINES [26], targeting compositions that retain a single-phase fcc structure over a wide temperature range. Among these, two alloys were identified as compositionally and structurally stable: a Ni-rich variant (HEA-1: $Cr_{14.2}Mn_{12.5}Fe_{39.7}Ni_{33.6}$) and a variant (Y3-HEA: : $Fe_{46.0}Mn_{17.0}Cr_{14.0}Ni_{23.0}$). The present study focuses on HEA-1, which was selected within the framework of the INNUMAT project for detailed atomistic and mechanical characterization [16]. The designation "HEA-1" is thus a project-specific shorthand used to distinguish this material from other experimentally studied Co-free HEAs, though only HEA-1 was examined at the MD level in this work. The study centers on the formation of deformation twins and their role in the mechanical response of polycrystalline HEAs. EBSD analyses provide spatially resolved insights into twin formation, showing strong agreement with atomistic features captured in MD simulations. Beyond twinning, the simulations also reveal the onset of dynamic recrystallization during deformation, offering additional understanding of the mechanisms governing high-temperature plasticity in these alloys. Together, these results deepen our knowledge of the thermally activated deformation processes in Co-free HEAs and support the design of compositionally complex materials optimized for reliable performance under extreme service conditions.

## 2. Methods

The studied alloy was cast by cold crucible induction melting using high-purity elemental precursors. To minimize chemical segregation and ensure compositional homogeneity, the ingot underwent homogenization annealing at 1200°C for 4 hours. The chemical composition was determined using X-ray fluorescence (XRF) for metallic elements and the inert gas fusion (IGA) method (LECO) for impurity analysis, as summarized in Table 1. Subsequently, the ingot (about 3 kg mass) was sectioned into multiple pieces (~450 g each) and hot forged into 12 mm diameter bars, achieving an approximate 80% reduction in cross-sectional area. The sample designated as IN70-1 was selected for this study. Finally, the material was recrystallized at 900°C for 1 hour, followed by air cooling. For consistency, the research alloy is referred to as HEA1 in the subsequent text.

Table 1. High entropy alloy composition and impurities.

| | Principal elements [wt. %] | | | | Impurities [wt. ppm] | | | |
|---|---|---|---|---|---|---|---|---|
| HEA-1 | Fe | Cr | Ni | Mn | C | S | O | N |
| | 39.7 | 14.2 | 33.6 | 12.5 | 9 | 35 | 8 | 5 |

## 2.1. Experimental methods

Specimens for structural observations were prepared with mirror-finished surfaces obtained by preliminary grinding using sandpapers up to a grit size of 1000x. Subsequently, electro-polishing procedure at 10°C, utilizing a mixture of ethanol and perchloric acid, was implemented. The microstructure of the produced materials was characterized using a Helios 5 UX scanning electron microscope (SEM) (ThermoFisher Scientific) equipped with an Electron Backscatter Diffraction (EBSD) system (EDAX Velocity Pro). EBSD data were acquired in the form of Inverse Pole Figure (IPF) orientation maps using a 20 keV electron beam and a step size of 0.5 µm. Post-processing analysis was performed using OIM Analysis 8 software. To improve data quality, a grain confidence index (CI) standardization procedure was applied to all datasets as a clean-up step, followed by removal of points with CI < 0.1. This procedure does not alter the crystallographic orientation of the grains but enhances the reliability of the measured microstructural parameters. Structural analysis was performed using a Bruker D8 Advance diffractometer with Cu Kα radiation ($\lambda$ = 1.5406 Å) in Bragg-Brentano geometry. Diffraction patterns were recorded in the 2θ range of 10°–145° at room temperature, utilizing a LYNXEYE XE-T detector operating in high-energy-resolution 1D mode with a 2.941° opening angle. Phase identification and model refinement were conducted using the Bruker DIFFRAC.TOPAS software, employing fundamental parameters profile fitting (FPPF) and Rietveld refinement, with reference to the JCPDS-ICDD PDF-4+ 2023 database.

Mechanical parameters were determined through uniaxial tests of miniaturized samples with a gauge length of 5 mm and a cross-section of 0.8 × 0.6 mm. Tensile experiments were performed at an initial strain rate of $10^{-3}$ $s^{-1}$ using a Zwick/Roell Z020 static testing machine equipped with a 20 kN load cell and a non-contact laser system for strain measurements. The 0.2% offset yield strength (YS) and tensile strength (TS) were calculated in accordance with the ISO 6892-1 standard. Tests were conducted at room temperature, 400°C, 550°C, and 700°C. At least four measurements were taken at each temperature, and the mean value with standard deviation was calculated.

## 2.2. Computational Methods

To perform our MD simulations, we use Large-scale Atomic/Molecular Massively Parallel Simulator (LAMMPS) [51] software which allows us to study the behavior of materials under a wide range of conditions. One of our goals is to accurately model plastic deformation, which is a crucial aspect of how materials respond to external loads. We employ a recently developed embedded atom method (EAM)-based interatomic potential that was specifically tailored for the plasticity-oriented behavior of the CrFeMnNi quaternary HEA [40,43]. The potential, derived through an extensive fitting process involving experimental data, DFT calculations, and thermodynamic modeling, accurately captures critical crystallographic properties such as elastic constants and stacking fault energy [18,19]; making it ideally suited for the work presented here. To create the fcc $Cr_{15.3}Mn_{12.8}Fe_{39.8}Ni_{32.1}$ at% sample for tensile test simulations, we begin by defining a numerical cell of pure Ni with a lattice constant of 0.359 nm along the main crystal orientations with Atomsk [52]: [001] with 261,000 atoms, [101] with 255,780 atoms, and [111] with 264,450 atoms, maintaining a density of 8.23 g/cm³, in good agreement with the experimental value. The fully periodic numerical cells have dimensions of approximately 11 nm in the x and y directions and 26 nm in the z direction. The initial Ni sample is then modified by substituting Ni atoms with 39.8% Fe, 12.8% Mn, and 15.3% Cr, in accordance with experimental results. Energy minimization is performed using the FIRE algorithm with a tolerance of $10^{-6}$ eV [24], ensuring that the system reaches its lowest-energy configuration [43]. The optimization criteria require that the energy change between successive iterations and the most recent energy magnitude remain below $10^{-5}$ eV. Additionally, the global force vector length for all atoms is constrained to be less than or equal to $10^{-8}$ eV/Å [49,50]. The samples are then equilibrated at room temperature, 400°C, 550°C, and 700°C for 2 ns using an isobaric-isothermal ensemble. This equilibration is achieved by integrating the Nose–Hoover equations with damping parameters: T = 2 fs for the thermostat and TP = 5 ps for the barostat, while maintaining an external pressure of 0 GPa [49,50].

To model the behavior of polycrystalline materials under mechanical loading, the initial configuration consisted of a pure Ni single crystal with approximately 1.55 million atoms in a simulation box measuring 20 × 21 × 40.5 $nm^3$. The polycrystalline structures were generated using the Voronoi tessellation method implemented in Atomsk [51]. In this approach, the position and crystallographic orientation of each grain were explicitly defined using the node keyword in the Atomsk input file. Each line of the input specified the spatial coordinates of the grain seed (the center of the Voronoi cell) and its crystal orientation, described by Euler angles ($\phi_1$, $\Phi$, $\phi_2$) in the Bunge convention. The grain orientations were obtained from experimental EBSD measurements of the $Cr_{14.2}Mn_{12.5}Fe_{39.7}Ni_{33.6}$ alloy, where Kikuchi diffraction patterns were indexed to determine local orientations and grains were reconstructed by grouping pixels with misorientations below 5°. The seed positions were chosen to fill the simulation box uniformly without introducing artificial periodicity, while the assigned orientations directly reflected the experimental EBSD dataset. Atomsk then performed the Voronoi tessellation to partition the domain into distinct grains, each populated atomistically according to its orientation. The resulting models contained up to 12 grains, corresponding to an effective grain size of 6–8 nm, which is typical for molecular dynamics simulations. The simulation box dimensions were identical to those used for the Ni single-crystal model, enabling direct comparison between single-crystal and polycrystalline systems. Although the absolute grain size in the MD models is significantly smaller than in the experimental specimens, ten independent polycrystalline configurations were generated using different random seeds to define the grain nuclei. This approach reproduces the statistical orientation and misorientation distributions observed experimentally, ensuring that the simulated microstructures provide an atomistically meaningful and statistically representative description of the alloy's polycrystalline behavior. To ensure the atomistic model is representative of the bulk polycrystalline behavior, we adopted the concept of a Representative Volume Element (RVE) being a simulation cell large enough to capture the essential features of grain interactions, such as dislocation-grain boundary interactions, grain boundary sliding, and misorientation effects, while remaining computationally tractable. Several polycrystalline configurations with different grain numbers (up to 12 grains) were generated and tested to evaluate the size effect on the stress–strain response. Convergence of mechanical behavior with respect to system size was used as a criterion to confirm the representativeness of the chosen models. While the absolute grain size in MD is typically smaller than in experimental specimens due to computational limits, the modeled RVEs provide statistically valid insight into local deformation mechanisms that scale to experimental observations. These RVEs were then used to study both pure Ni and high-entropy alloy (HEA) polycrystals, prepared in the same way as their corresponding single-crystal counterparts and equilibrated across a range of temperatures prior to mechanical testing in MD. The differences in absolute yield strength between experiments and MD simulations can be largely attributed to the intrinsic grain-size scaling dictated by the Hall–Petch relation. The polycrystalline representative volume elements (RVEs) employed in MD contain only a few grains with dimensions of a few tens of nanometers, several orders of magnitude smaller than those in the experimental samples. At this scale, grain boundaries act as strong barriers to dislocation motion, leading to an artificial strengthening effect that elevates the simulated yield stress. This nanoscale strengthening does not affect the qualitative trends of the stress–strain response but shifts the absolute values upward relative to macroscopic measurements. Therefore, the MD results should be interpreted as providing atomistic insight into deformation mechanisms, such as dislocation-grain-boundary interactions and twin nucleation, rather than direct quantitative predictions of bulk strength. The agreement between the temperature-dependent trends observed in simulations and experiments nonetheless confirms that the modeled RVEs capture the essential physics governing the mechanical response of HEA-1.

To simulate a tensile test on the material, with a strain rate of $10^9$ $s^{-1}$. In each computational step, the constituent particles within the atomistic ensemble were remapped to the instantaneous dimensions of the simulation cell. This ensured displacement-controlled straining of the cuboidal simulation cell along the straining direction (the z-axis), while a barostat was applied in the remaining two directions (x and y) to maintain constant pressure in both. The stress during deformation was evaluated directly from the virial stress tensor calculated in the MD simulations and normalized by the instantaneous simulation-cell volume, thereby accounting for both interatomic forces and atomic kinetic contributions. The system temperature and pressure were regulated using a Nose–Hoover thermostat (damping coefficient 0.1 ps) and a Parrinello–Rahman barostat (damping coefficient 1 ps). Tensile deformation was applied along the z-axis, while the barostat operated only in the x and y directions to maintain zero lateral pressure ($P_x = P_\gamma = 0$), ensuring a uniaxial stress state consistent with the applied strain. The uniaxial strain imposed on the system was calculated using the following relation:

$$\varepsilon = \frac{l_0 - l_z(t)}{l_0} \qquad (1)$$

where $L_0$ represents the initial cell length along the z-axis before strain was applied, and $L_z(t)$ is the instantaneous cell length along the z-axis at time t. The time step for the simulation was set to 2 fs to ensure the accuracy and stability of the results throughout the deformation process. For each temperature and sample, we perform 10 MD simulations by considering different seeds to prepare the HEA-1s to accomplish for statistics in our results. Finally, The Young's modulus was calculated from the linear elastic region of the stress–strain curve by performing a linear least-squares fit within the strain range of $0 \leq \varepsilon \leq 0.01$, which corresponds to the purely elastic regime prior to dislocation nucleation.

To understand the plastic deformation mechanisms of twin formation during tensile test, we consider that the Generalized Stacking Fault Energy (GSFE) is a critical parameter able to quantify the energy variation associated with dislocation motion through the crystal lattice. The GSFE calculations were performed for the {111} slip plane of the fcc lattice, applying periodic boundary conditions along the cut plane. Along the z-axis, which is normal to the (111) slip plane, shrink-wrapped boundary conditions were applied to create two free surfaces at the top and bottom of the slab. This configuration allows full relaxation of the lattice in the normal direction, ensuring that the normal stress components ($\sigma xz$, $\sigma yz$, $\sigma zz$) remain effectively zero throughout the imposed shear process. The top and bottom halves of the simulation cell were incrementally displaced along the [110] shear direction, corresponding to multiples of the Burgers vector, while the atoms were allowed to fully relax perpendicular to the fault plane (along the [111] direction). This relaxation ensures accurate minimization of interplanar spacing and local atomic distortions during shearing, using a replicated sample from an fcc unit cell with 13957 Ni atoms and a dimension of 4.31x2.59x14.02 $nm^3$; for the fcc $Cr_{14.2}Mn_{12.5}Fe_{39.7}Ni_{33.6}$, we followed the process explained before. Displacements are applied in equal increments, each representing 0.1 of the Burgers vector magnitude. Following each displacement, the top and bottom atomic layers are fixed, while the remaining layers relax exclusively in the $y$ direction. This relaxation process is performed using energy minimization with the conjugate gradient method and is considered complete when either (i) the ratio of the energy change between successive iterations to the most recent energy magnitude is less than $10^{-12}$, or (ii) the global force vector magnitude for all atoms is less than or equal to $10^{-12}$ eV/Å. Subsequently, the stacking fault energy can be calculated as: $\gamma GSFE = \frac{Es-E0}{ASF}$ where $E_s$ represents the energy of the sample at a given displacement, and $E_0$ denotes the energy for the perfect sample, $A_{SF}$ stands for the stacking fault area.

## 3. Results

As the first step of the investigation, the structure of the HEA-1 alloy was analyzed. Figure 1 presents EBSD maps of the alloy in two states: the as-delivered (as forged) condition and after the recrystallization process (900°C/1h). In the as-delivered state, the material exhibits a significantly deformed structure with fine grains (Fig. 1(a)). The microstructure is characterized by a refined grain size and pronounced intragranular deformation, with the deformation process generating new grain boundary interfaces; twin boundaries remain scarce, contributing less than 1% to the total grain boundary area. The recrystallization process, followed by grain growth, leads to the formation of new deformed equiaxed grains. Their size depends on the recrystallisation conditions [26]. This grain growth is accompanied by a reduction in the overall grain boundary volume (Fig. 1(b)). In the studied case, due to recrystallization, the average grain size increases by an order of magnitude, while the fraction of twin boundaries rises to nearly 40% of the total grain boundary area, determined by the analysis of the presented IPF map.

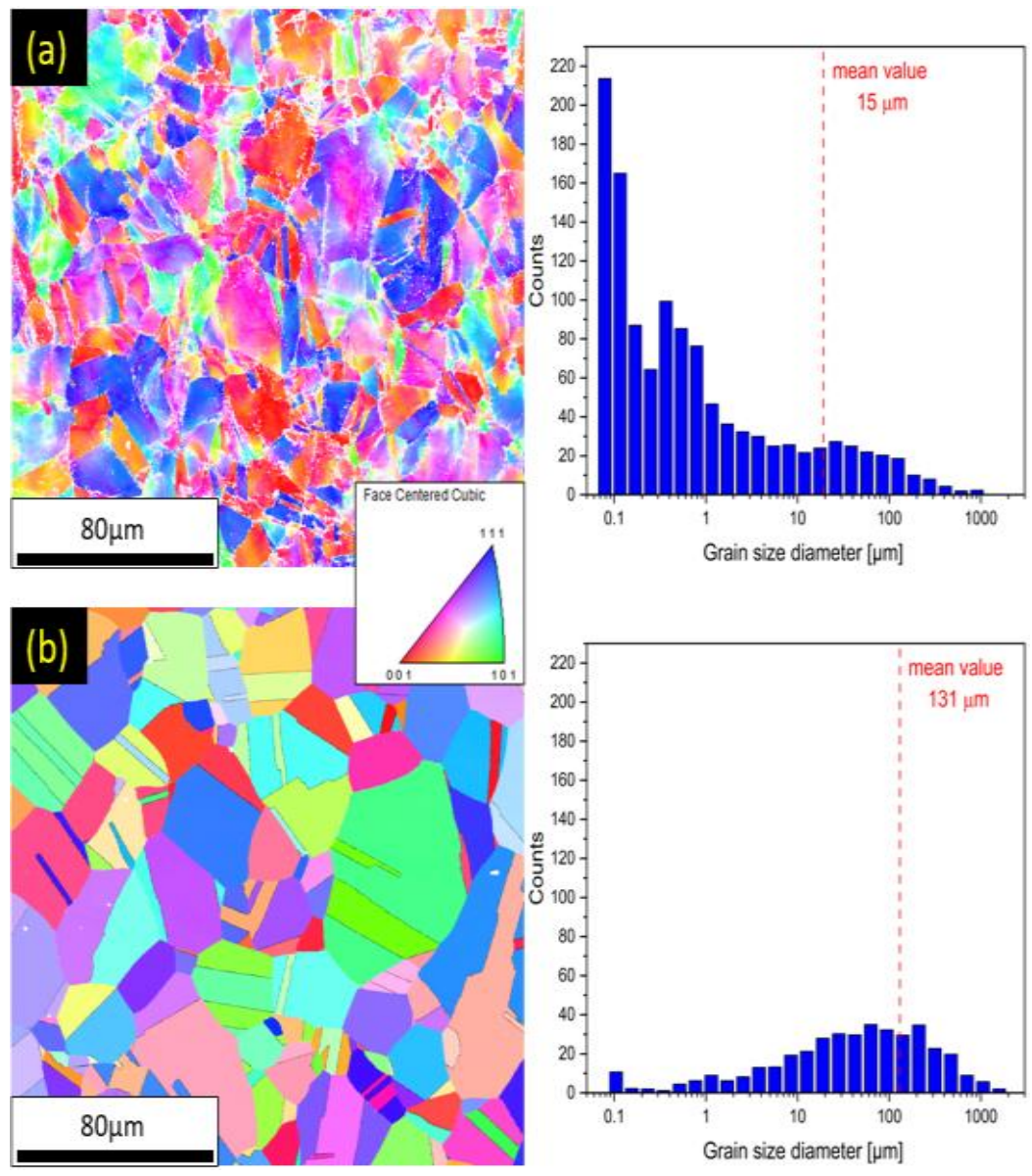

Figure 1. EBSD maps and corresponding grain size of HEA-1 alloy (a) in a as forged condition, and (b) as recrystallized condition

XRD measurements performed on the initial sample at room temperature (25°C) to establish a reference microstructure and phase composition prior to mechanical testing and guidance to the MD modeling which is obtained by the diffraction algorithm implemented into LAMMPS [52] (Fig. 2). Panel (a-b) compares the XRD patterns of the experimental and simulated samples reaching a good agreement for the lattice constant of 0.359 Å. In the MD results, virtual selected area electron diffraction (SAED) patterns are generated by examining the region of reciprocal space that intersects the Ewald sphere, which has a radius of $\lambda^{-1}$[53]. The diffraction peaks confirm the fcc crystal structure of HEA-1 and exhibit strong agreement between the experimental data and the computational model; the consistency in peak positions and relative intensities indicates that the simulated structure accurately reproduces the atomic arrangements and phase stability, while slight differences in peak intensities are attributed to the presence of minor texture in the sample. The experimental data is in good agreement with previously reported XRD spectrum for Co-free alloys [61]. Panel (c) illustrates the schematic representation of the MD polycrystalline model employed for tensile simulations, showing a cross-section of the sample. The model captures the essential features of the experimental microstructure, including grain morphology and boundary distributions, to accurately simulate the mechanical response under tensile loading.

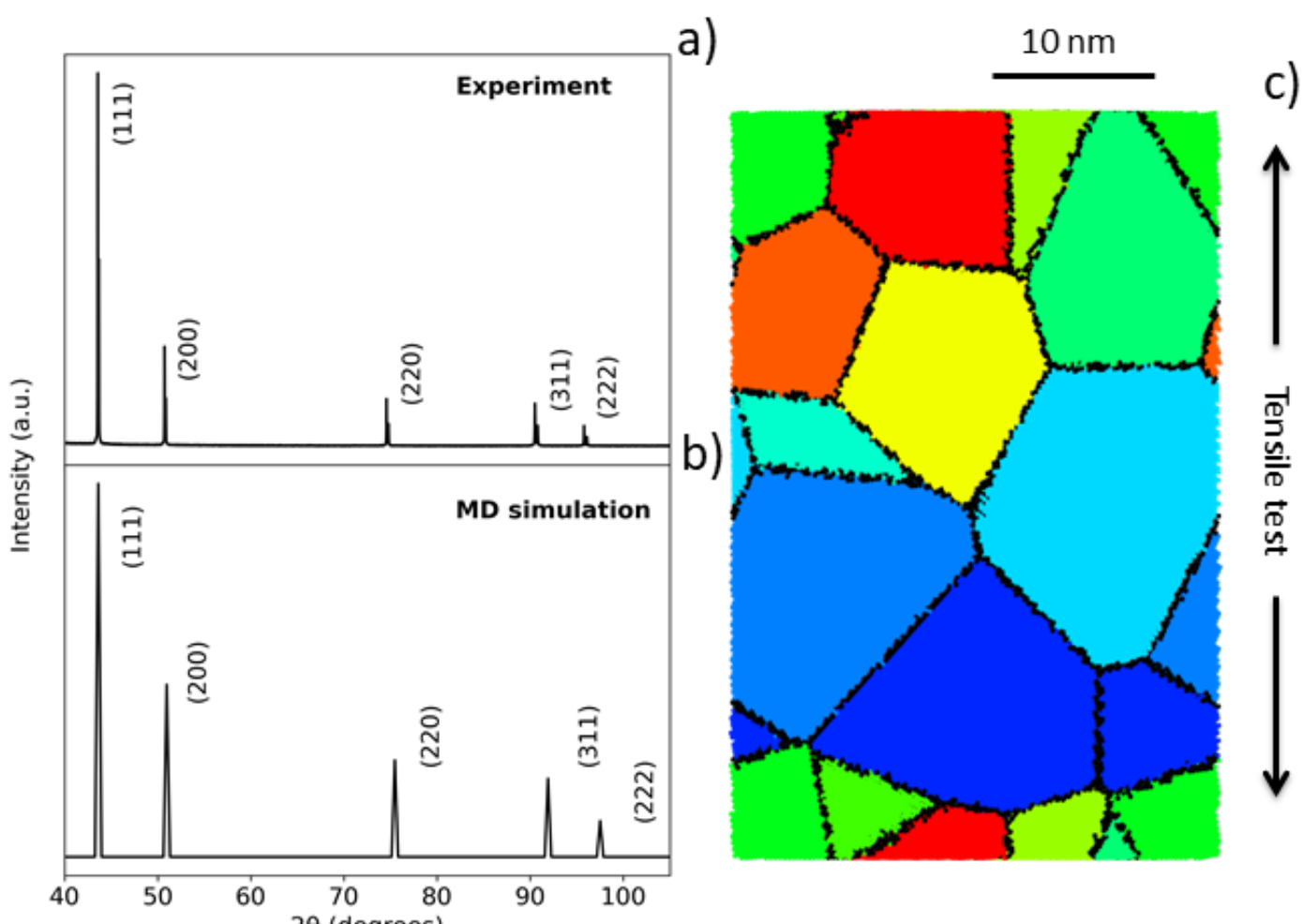

Figure 2. XRD patterns of the experimental and simulated samples at room temperature in recrystalized condition, demonstrating good agreement (a, b), and schematic representation of the polycrystalline MD model used for tensile simulations (c) color palette follow the IPF EBSD one shown in Fig 1.

### 3.1. Nickel effects on plasticity mechanisms of CrMnFeNi alloy

In Fig. 3, we present the computed GSFE for various displacement vectors, with atomic positions relaxed only perpendicular to the cut plane. The γ-lines along the <112>/2 (panel a) and <110>/2 (panel b) directions on the {111} plane are shown. We observed that the GSFE results for HEA-1 are lower than those for pure Ni samples, where an unstable stacking fault occurs along the <110>/2 direction at half of the Burgers vector (b), demonstrating the fcc nature of both materials. In contrast, along the <112>/2 slip direction, a stable stacking fault forms at b/3 (where b =<112>/2), and an unstable stacking fault occurs at b/6, with the latter being more pronounced for HEA-1 due to lattice mismatch [23,24]. The GSFE curve reveals the energy barriers that dislocations encounter as they traverse different stacking sequences in the crystal, including the ideal lattice structure and local stacking faults in good agreement with reported ab initio results [28].

In fcc HEAs, the GSFE plays a crucial role in governing dislocation glides and overall plastic deformation behavior. A lower SFE facilitates the activation of multiple slip systems and deformation twinning, enhancing ductility, while a higher GSFE restricts dislocation motion, promoting work hardening and increasing strength. The GSFE is highly sensitive to alloy composition, as variations in atomic size, local chemical interactions, and lattice distortions influence dislocation dynamics, ultimately tailoring the mechanical properties of the alloy. A key advantage of HEAs arises from the interplay between low SFE, which fluctuates with alloying composition, and high friction stress, leading to an exceptional balance between strength and ductility [23,24,25, 39,40] In Ni, a relatively high SFE results in a narrow dislocation dissociation width and lower friction stress, facilitating easier dislocation movement. In contrast, HEA-1, characterized by a lower SFE, exhibits a wider dissociation width, which enhances dislocation pinning and impedes dislocation glide [43, 47]. This strong pinning effect contributes to the alloy's superior strain hardening capacity and mechanical stability. The competition between slip and twinning mechanisms is further analyzed using the twinnability parameter ($\tau_a$), which combines information from the intrinsic, unstable, and twinning fault energies obtained from the GSFE curves as follows [69,70,71]:

$$\tau_{\alpha} = \left[1.136 - 0.151 \frac{\gamma_{\mathrm{ISF}}}{\gamma_{\mathrm{USF}}}\right] \sqrt{\frac{\gamma_{\mathrm{USF}}}{\gamma_{\mathrm{UT}}}} \tag{2}$$

where $\gamma$ISF, $\gamma$USF, and $\gamma$UT represent the intrinsic, unstable stacking-fault, and unstable twinning-fault energies, respectively. The calculated values are $\tau_a = 1.0$ for pure Ni and $\tau_a = 1.22$ for HEA-1, where results for Ni are in close agreement with previously reported data for fcc metals and the cantor alloy with a factor of 1.26 [71]. The marginal increase in $\tau_a$ for HEA-1 reflects the effect of compositional complexity and its slightly higher stacking-fault energy, implying a somewhat lower twinning propensity compared with pure Ni. Nevertheless, the similarity of the two values suggests that both systems remain within the regime where twinning can coexist with dislocation glide during plastic deformation, consistent with the experimental and atomistic observations presented in this work.

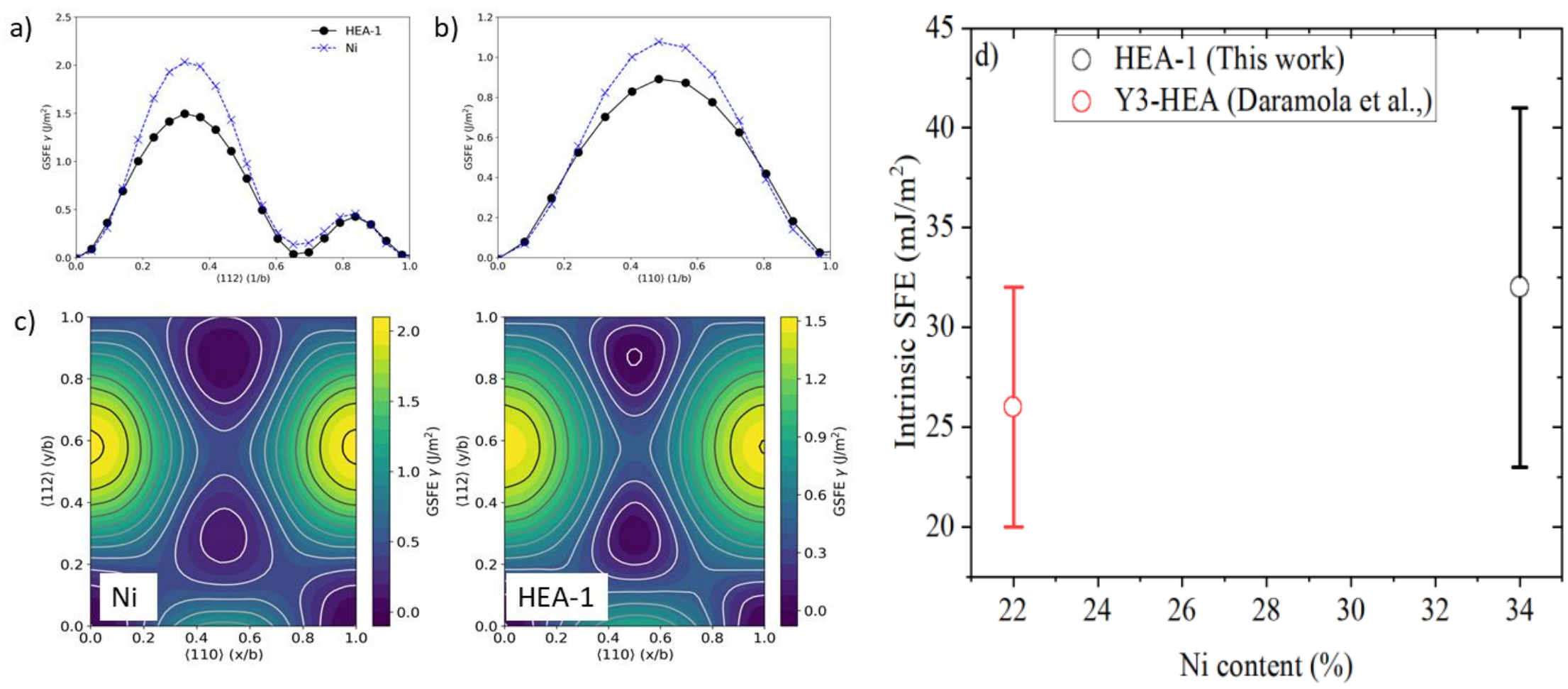

Figure 3. γ-lines along the (a) 〈112〉/2 and (b) 〈110〉/2 directions. (c) γ-surface on the loosest packing {111} planes for the pure Ni and HEA-1 samples. (d) Intrinsic stacking fault energy (SFE) estimated using the EAM potential for HEA-1 and Y3-HEA, based on calculation from Daramola et al. [15]. The error bars represent the standard deviation of SFE values across 6000 different stacking fault configurations.

The investigated alloy HEA-1 closely aligns with the composition of the Y3-HEA (Cr15Fe46Mn17Ni22) [25, 26] where intrinsic stacking fault energy (SFE) calculations for these alloys illustrate this trend, as shown in Fig. 10d. We estimated an SFE of $32 \pm 9$ mJ/m² for HEA-1, whereas Y3-HEA with its lower Ni content exhibits a slightly lower SFE of $26 \pm 6$ mJ/m². This increase in SFE with increasing Ni content is consistent with Liu et al.'s findings for CoCrFeMnNi alloys [35]. As a benchmark, we compute the ISFE of other fcc systems, for pure Ni, the computed value of $125 \pm 10$ mJ m$^{-2}$ agrees closely with the experimental range of 120–130 mJ m$^{-2}$[65], confirming that the potential reliably reproduces the energetics of faulted configurations in elemental fcc metals. To further assess its transferability to multicomponent alloys, we examined a 316-type austenitic stainless steel, which nominally contains Fe–17Cr–12Ni–2.5Mo–1Mn–1Si–0.05C (wt%). Because the available potential is parameterized for the Fe–Ni–Cr subsystem, the alloy was modeled as a Ni–Fe–Cr ternary solid solution ($Fe_{70}Cr_{18}Ni_{12}$ at%), representing the dominant chemistry of 316 SS. The calculated $\gamma$_ISF for this model alloy is $30 \pm 5$ mJ m$^{-2}$, which falls within the lower bound of experimentally reported values of 25–35 mJ m$^{-2}$ [66] and is comparable to that obtained for the HEA-1. This similarity reflects the Fe-rich character of both systems, which leads to a reduced fault energy relative to Ni-rich compositions. Consequently, the predicted differences in $\gamma$_ISF between the Ni-rich HEA-1 and Fe-rich Y3-HEA can be interpreted as physically meaningful and consistent with known alloying trends, providing a sound basis for analyzing their distinct deformation mechanisms and twinning behavior.

Furthermore, tensile tests at room temperature and a strain rate of $10^{-3}$ s$^{-1}$ by Traversier et al. [25, 26] report a YS of $210 \pm 10$ MPa for Y3-HEA, while Fig. 3 shows HEA-1 at $220 \pm 13$ MPa. Consistent with the defect landscape, HEA-1 exhibits a higher yield strength (220 Mpa; Tab. 2) than Y3-HEA (210 MPa) [26], indicative of increased dislocation obstacles, while UTS values remain comparable within uncertainty [26]. A plausible explanation is that HEA-1 higher local SFE fluctuations (as indicated by its error bars) will produce higher lattice stresses that will effectively impede dislocation motion more than Y3-HEA. This result is consistent with our experimental observations and those reported by Smith et al. [36] on Cantor-alloy–type systems, suggesting that complex solute arrangements and the resulting local SFE fluctuations drive the observed variations in dislocation dissociation width in concentrated alloys. In addition, the SFEs we measured for both HEA-1 and Y3-HEA (26–32 mJ/m²) lie within the typical 26–38 mJ/m² range reported for the Cantor alloy [25-27]. Such low SFEs enhance the propensity for deformation twinning and ε-martensitic transformation: with γ-fcc and ε-HCP phases of nearly equal stability, local stress or temperature fluctuations can readily shift the phase balance. In fcc metals, SFEs above ~15 mJ/m² tend to favor planar dislocation slip and twinning, whereas lower SFEs promote dissociation into partials separated by stacking faults. These observations underscore the critical roles of Ni and Co in tuning SFE—and thus controlling the alloy's deformation mechanisms and mechanical response.

The temperature dependence of the SFE in HEA-1 was determined using a finite-temperature free-energy approach [67,68]. In this method, both perfect and faulted configurations were equilibrated in the NVT ensemble at each target temperature (25, 400, 550, and 700 °C). The shear displacement along the {111}⟨112⟩ slip path was treated as a collective variable, and the mean constraint force associated with maintaining each displacement was computed from MD runs. Integration of this mean force along the displacement coordinate yielded the potential of mean force G(s, T), from which the intrinsic stacking-fault free energy was computed as: (G(s,T)-G(0,T))/ASF. This approach naturally incorporates thermal expansion and vibrational entropy, providing a physically meaningful description of stacking-fault energetics under finite-temperature conditions. A gradual decrease in stacking-fault energy with increasing temperature was observed, from $32 \pm 9$ mJ m$^{-2}$ at 25 °C to $22 \pm 6$ mJ m$^{-2}$ at 700 °C. This trend agrees with the temperature dependence reported for fcc Cu [68] and arises from the increasing entropy contribution, which lowers the free-energy difference between faulted and perfect configurations. The reduced ISFE at intermediate temperatures (400–550 °C) promotes partial dislocation dissociation and twin formation, consistent with the experimentally observed increase in twin boundary density in this range. At higher temperatures, however, additional recovery processes and grain-boundary sliding become dominant, leading to a suppression of twinning and a transition toward more homogeneous deformation.

Finally, for the non-equiatomic fcc $Cr_{15.3}Mn_{12.8}Fe_{39.8}Ni_{32.1}$ alloy, which is equilibrated at room temperature, 400 °C, 500 °C, and 700 °C through computational methods, it is assumed that the atomic configurations are random in both single-crystal and polycrystalline models. Although this alloy does not exhibit complete chemical disorder, instead, pronounced short-range order (SRO) tendencies have been reported [23,24], particularly between Ni and Mn atoms. This SRO behavior may promote the formation of $L1_0$-like local structures, which are known to act as precursors for deformation mechanisms such as twinning, as also evidenced in experimental observations. Ongoing work will perform more detailed SFE calculations to confirm these estimates and refine the comparisons [27, 28]. Recent studies have also linked intrinsic SFE to irradiation behavior in materials similar to HEA-1 [43, 58-60]; accordingly, we will next investigate how SFE influences the irradiation response of HEA-1.

### 3.2. Mechanical response

Figure 4a) shows experimental stress-strain curves for HEA-1 at different temperatures obtained from tensile testing. It is observed that the yield point decreases as the temperature increases, indicating that the material transitions from elastic to plastic deformation at lower stress levels with higher temperatures. As strain increases, the material enters the strain-hardening region, where the stress continues to rise, suggesting strengthening through dislocation interactions and twin formation. The curve peaks at the ultimate tensile strength (UTS) of 500 MPa at room temperature, but the UTS decreases significantly with increasing temperature, reflecting the maximum stress the material can withstand before necking occurs [25,26]. After this point, stress declines as necking initiates, leading to localized deformation. Finally, the material fractures at a strain of 55% at room temperature, with less strain required for fracture at elevated temperatures. In Fig 4b), we present the calculated yield strength (YS) and UTS as a function of the temperature for reference.

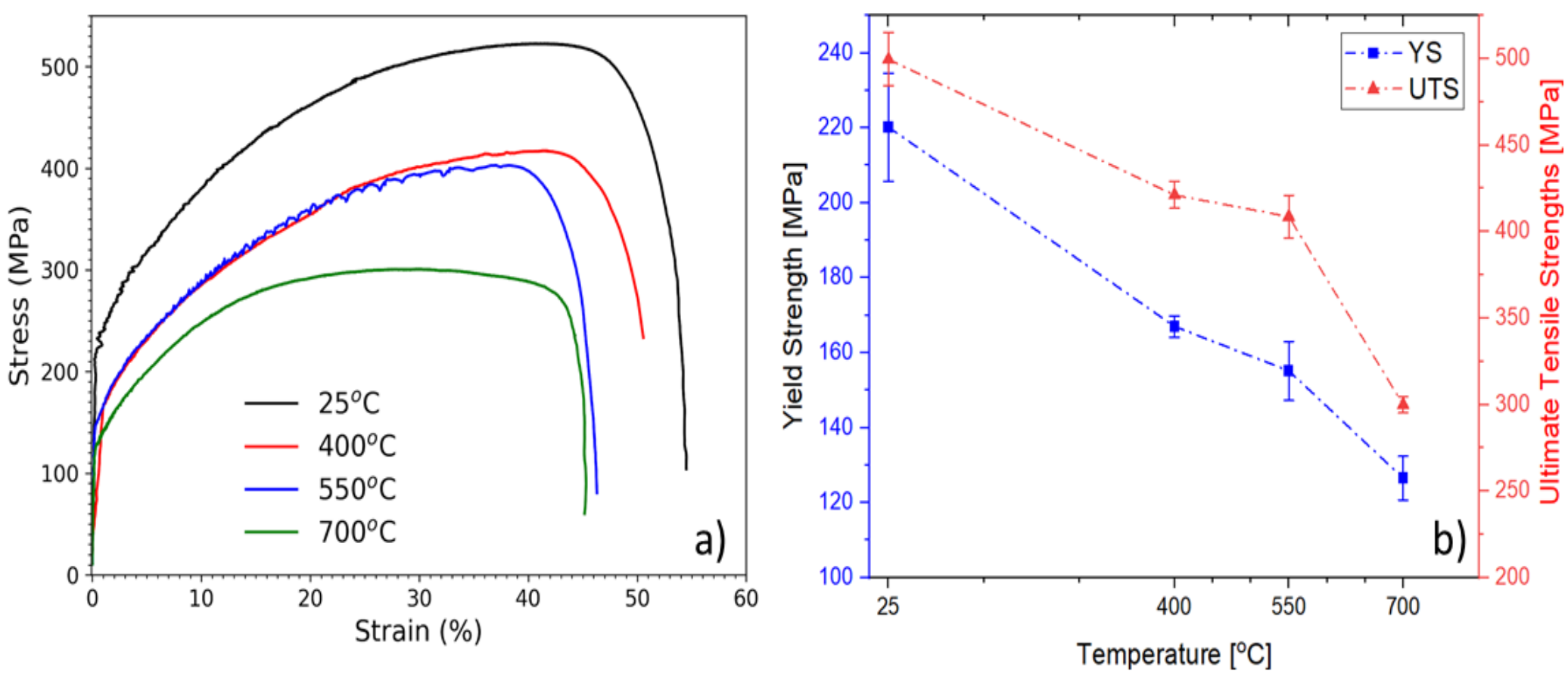

Figure 4. Stress-strain curves for polycrystal HEA-1 as a function of the applied strain for each temperature for experimental results in a) and yield and ultimate strengths as a function of the temperature in b)

Fig 5 shows results for the MD simulations where the linear increase in the elastic region, where the Young's modulus is observed to decrease with increasing temperature. The yield point is reached at different strain values depending on the temperature. Beyond a strain of 0.04, HEA-1 enters the plastic deformation regime at all temperatures. Grains with slip-favorable orientations (e.g., [111]) yield earlier and actively accommodate plastic strain, whereas grains with less favorable orientations (e.g., [001]) require higher applied strain to initiate plastic deformation. These results qualitatively agree with the experimental data, where the sequence of stress-strain behavior at different temperatures is followed, and the reduction in both yield point and Young's modulus is well captured, alongside the decrease in strain to fracture at higher temperatures.

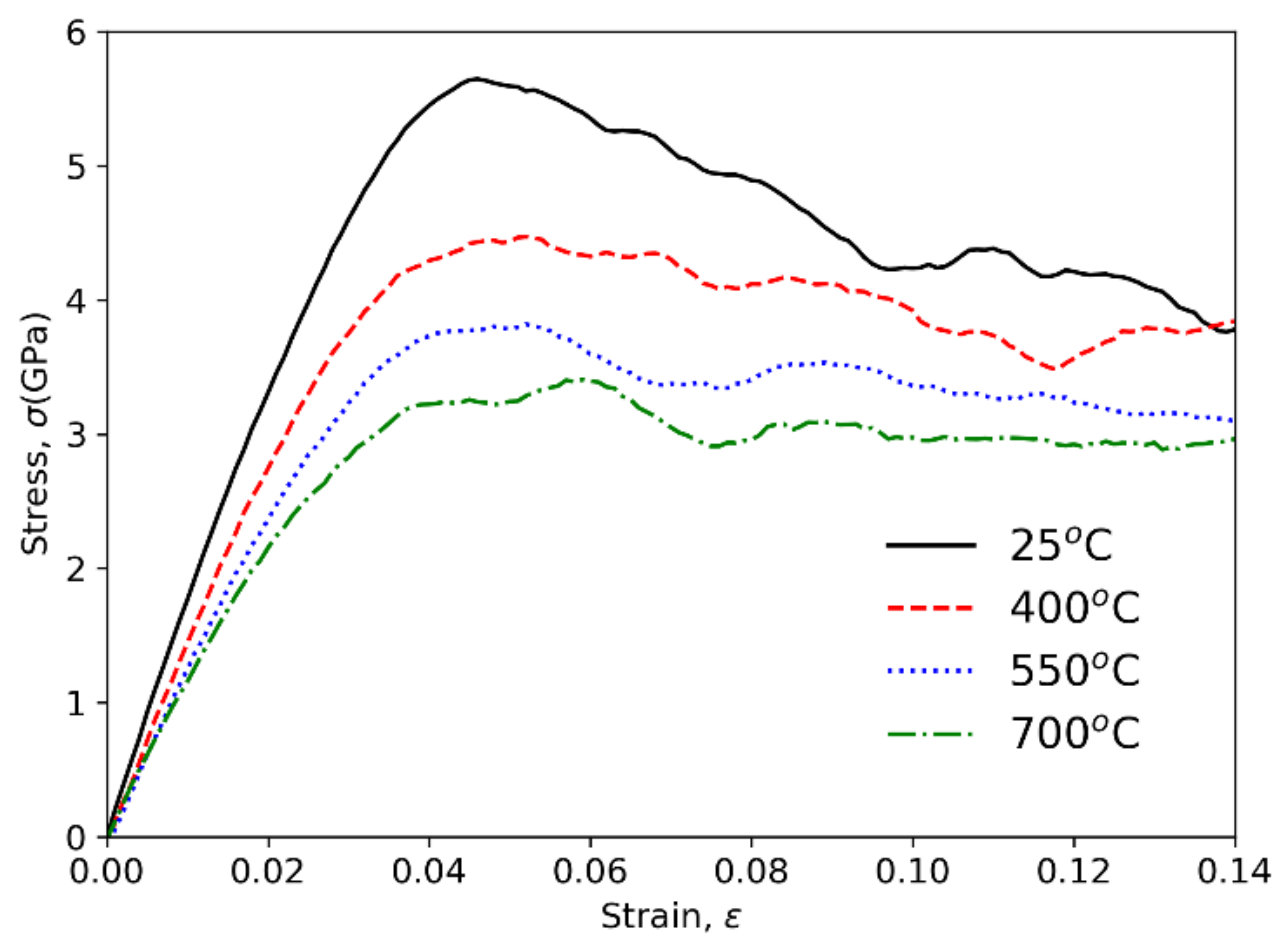

Figure 5. MD simulation results showing stress–strain curves as a function of temperature for (a) a single crystal and (b) a polycrystalline HEA-1 alloy.

To compare experimental data with MD results, the YS was determined using the 0.2% offset method, as done by experimental analysis. The elastic region of the stress-strain curve was fitted using the linear relationship $\sigma = Y_{MD}\,\epsilon$, where $Y_{MD}$ is the elastic modulus obtained from the initial slope. The yield point was then defined as the intersection of the stress-strain curve with a line offset by 0.002 strain from the linear fit. The UTS was obtained by identifying the maximum stress value in the stress-strain curve. Using these experimental and MD-derived results, the normalized UTS was computed as: $\chi$ = UTS(T)/UTS(25C), along with a normalized yield strength (YS) as a function of temperature. Figure 6 presents (a) the normalized UTS and (b) the normalized YS. We observed that MD results are in good agreement with experimental measurements. Moreover, Co is one of the primary elements in many HEA alloys, providing fcc stabilization, outstanding mechanical performance, especially at low and cryogenic temperatures, and favorable plasticity. Nevertheless, due to its high cost and radioactivity in nuclear environments, novel Co-free HEA materials are being developed that can deliver a comparable set of properties. In Figure 5, alongside our work, we have also included reported results for CoCrFeMn, CoFeNiMn, and the Cantor alloy [1–4], comprising both Co-containing and Co-free materials, highlighting the influence of chemical complexity on mechanical performance. Among these, the CoCrFeMn alloy exhibits the lowest UTS, while the Co-free HEA-1 demonstrates the highest UTS, underscoring the impact of composition on strengthening mechanisms. In addition, Figure 5, along with comparisons to literature data, demonstrates that HEA-1 exhibits mechanical properties that in some cases surpass those of Co-containing alloys, particularly up to 550 °C; a noteworthy achievement [23-27]. This result highlights the effectiveness of chemical and structural optimization strategies in Co-free systems, confirming that targeted compositional design can successfully enhance high-temperature mechanical performance. The grain size of the polycrystalline models used in the MD simulations is considerably smaller than that of the experimental samples, which can influence the apparent yield behavior. At such nanometric scales, deformation is dominated by grain-boundary-mediated processes, including dislocation emission, sliding, and migration, rather than by traditional dislocation storage and propagation observed in larger grains. As a result, the atomistic models tend to exhibit higher yield stresses but capture the same underlying deformation modes and trends with temperature. Therefore, the MD polycrystals should be interpreted as representative of the intrinsic local response of the alloy, providing valuable mechanistic insight even though the absolute stress levels differ from the macroscopic measurements [43, 72-74].

In addition, we provide a quantitative comparison of the absolute mechanical properties, the ultimate tensile strength (UTS) and yield strength (YS) values at different temperatures are summarized in Table 1 and Table 2, respectively. These data complement the normalized trends shown in Fig. 5, enabling a direct comparison between the Co-free HEA-1 and Co-containing reference alloys. The results for single-crystal simulations with various crystallographic orientations are presented in the Supplementary Material.

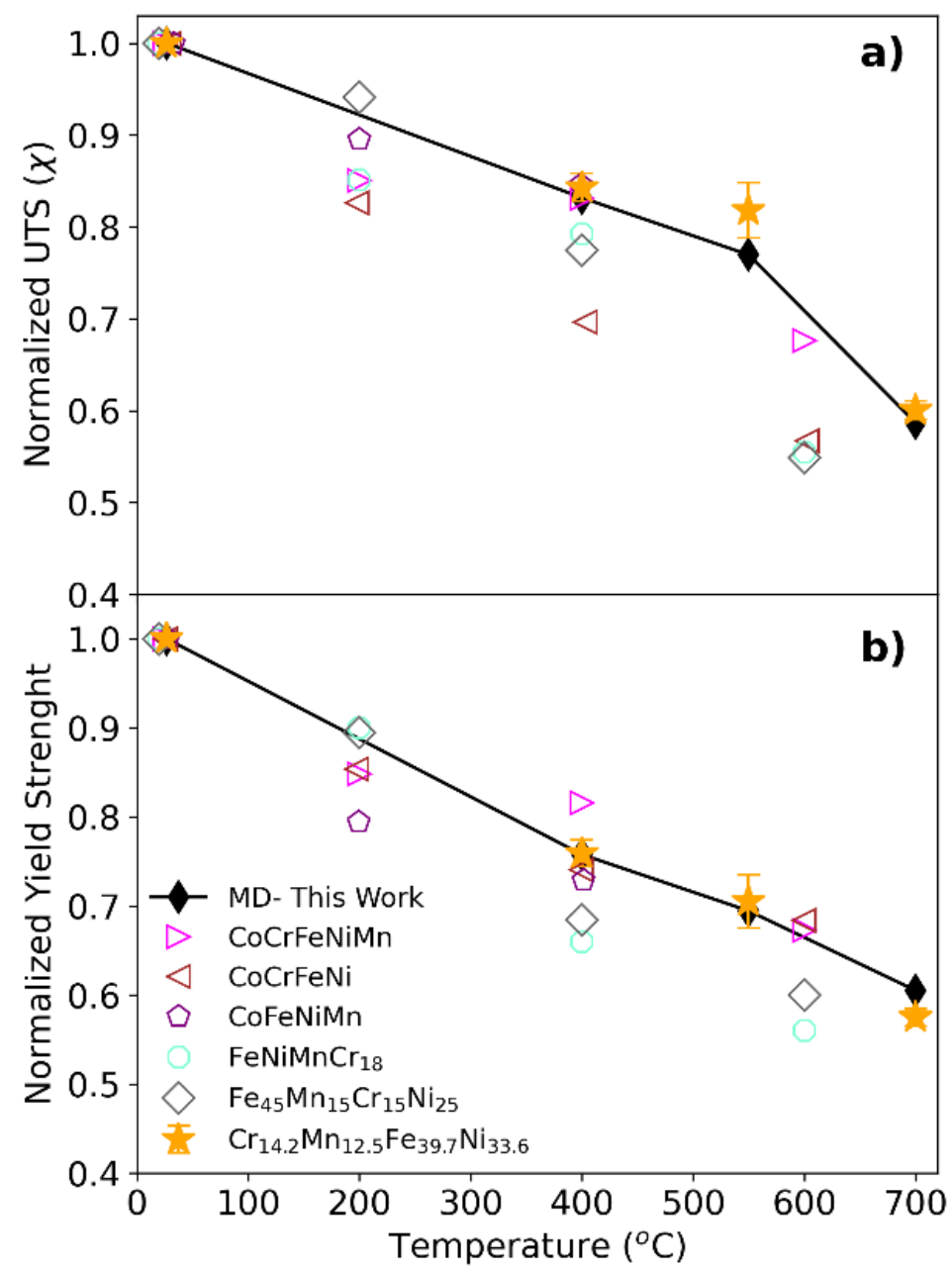

Figure 6. Comparison between experimental results and MD simulations for normalized UTS and yield strength (YS), demonstrating good agreement across the analyzed temperature range. We include results for the CoCrFeNi [1], CoFeNiMn [2], Cantor alloy for comparison [3,4], $NiFeCrMn_{18}$ [54], and $Fe_{45}Mn_{15}Cr_{15}Ni_{25}$ [55].

Table 1. Ultimate tensile strength (UTS) of the studied alloys at different temperatures obtained from experimental data and compared to CoCrFeNi [1], CoFeNiMn [2], Cantor alloy for comparison [3,4], $NiFeCrMn_{18}$ [54], and $Fe_{45}Mn_{15}Cr_{15}Ni_{25}$ [55]. The values are used to derive the normalized UTS trends presented in Fig. 5(a).

| Temp./Alloy | $Cr_{14.2}Mn_{12.5}Fe_{39.7}Ni_{33.6}$ | CoCrFeNi | CoFeNiMn | CoFeNiCrMn | $NiFeCrMn_{18}$ | $Fe_{45}Mn_{15}Cr_{15}Ni_{25}$ |
|---|---|---|---|---|---|---|
| 25 | 499.64 | 692.91 | 560.65 | 590.64 | 631 | 510 |
| 200 | - | 572.24 | 501.89 | 502.29 | 537 | 480 |
| 400 | 421.17 | 482.60 | 474.15 | 490.84 | 500 | 395 |
| 550 | 408.68 | - | - | - | - | - |
| 600 | - | 392.94 | - | 399.47 | 350 | 280 |
| 700 | 299.89 | - | - | - | - | - |
| 800 | - | 317.35 | - | 183.63 | - | - |

Table 2. Yield strength (YS) of the studied alloys at different temperatures obtained from experimental data and compared to CoCrFeNi [1], CoFeNiMn [2], Cantor alloy for comparison [3,4], $NiFeCrMn_{18}$ [54], and $Fe_{45}Mn_{15}Cr_{15}Ni_{25}$ [55]. The values are used to derive the normalized UTS trends presented in Fig. 5(a).

| Temp./Alloy | $Cr_{14.2}Mn_{12.5}Fe_{39.7}Ni_{33.6}$ | CoCrFeNi | CoFeNiMn | CoFeNiCrMn | $NiFeCrMn_{18}$ | $Fe_{45}Mn_{15}Cr_{15}Ni_{25}$ |
|---|---|---|---|---|---|---|
| 25 | 220.17 | 250.57 | 214.47 | 210.77 | 250 | 190 |
| 200 | - | 213.95 | 141.38 | 178.79 | 225 | 170 |
| 400 | 167.00 | 185.64 | 129.82 | 172.01 | 165 | 130 |
| 550 | 155.16 | - | - | - | - | - |
| 600 | - | 171.28 | - | 141.75 | 140 | 114 |
| 700 | 126.51 | - | - | - | - | - |
| 800 | - | 70.42 | - | 88.71 | - | - |

### 3.3. Structural analysis after tensile test

The activation of deformation twins provides an additional strain accommodation pathway, increasing the strain-hardening rate by introducing barriers to dislocation motion. However, the sudden formation and rearrangement of twins can lead to localized stress relaxation, manifesting as discrete drops in the stress-strain curve [35,36]. This effect is particularly pronounced in polycrystalline samples, where grain orientation plays a crucial role in twinning activity. Figure 7 presents EBSD orientation maps obtained from the rupture region of tensile specimens tested at different temperatures. The maps reveal variations in crystallographic orientation within individual grains, while grain boundaries are distinguishable by abrupt color changes. Numerous twin boundaries, marked by black lines, are clearly observed, and their fraction of the total grain boundary network depends on the test temperature. The crystallographic orientation relative to the surface normal is represented using color coding: red denotes the [001] orientation, blue corresponds to [111], and green indicates [101] (see inset of IPF map legend in Fig. 7b). At high temperatures and elevated strain levels, the specimen undergoes significant lattice distortion due to the accumulation of crystal defects. Variations in crystallographic orientation within individual grains arise from dislocations generated during plastic tensile deformation, which organize into diffuse networks and tangles. The evolution of intragranular orientation spread during deformation was quantified from the atomistic data by computing the misorientation angle between each atom’s local quaternion and the mean grain orientation. The resulting histograms as presented in the supplementary material reveal a progressive broadening of the orientation distribution with increasing strain, reflecting the accumulation of stored dislocations and local lattice rotations within the grains. This increase in orientation dispersion quantitatively supports the interpretation that, as plastic deformation advances, the density of geometrically necessary dislocations and localized shear regions rises, leading to a more heterogeneous crystallographic orientation field consistent with the experimental EBSD observations.

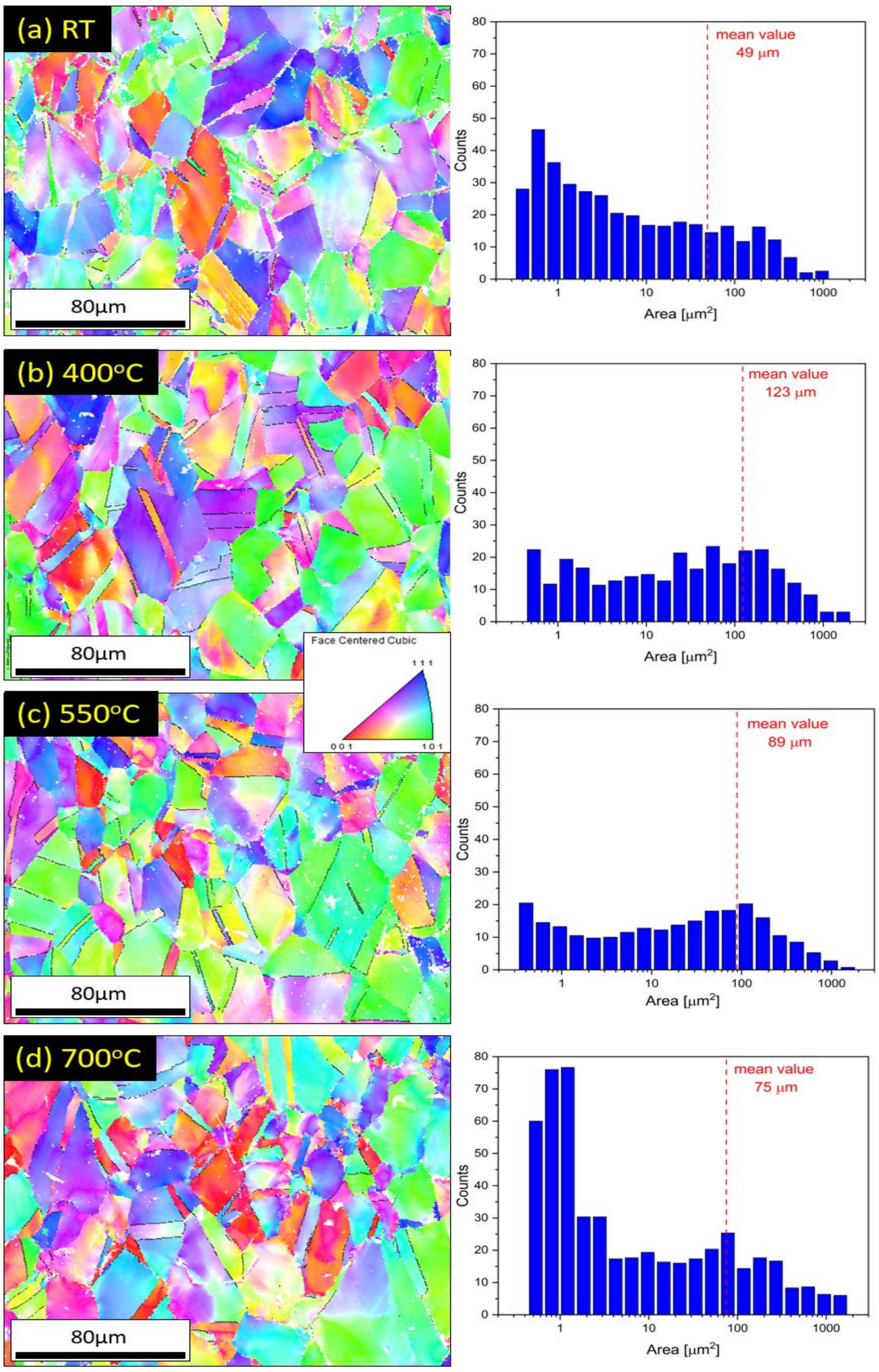

Figure. 7. EBSD orientation maps and corresponding grain size histograms obtained on specimens after tensile tests at different temperatures: (a) room temperature, (b) 400°C, (c) 550°C, and (d) 700°C.

The quality of EBSD patterns deteriorates with the accumulation of lattice defects, leading to a decrease in indexing reliability [56,57]. The presence of white regions in the IPF maps corresponds to mis indexed pixels with confidence index lower than 0.1, which can be attributed to localized lattice distortions. These distortions are particularly pronounced near GBs at room temperature (Fig. 7a), where high-stress concentrations intensify defect accumulation and hinder accurate orientation mapping. In addition, GBs were classified based on their misorientation angles. Low-angle grain boundaries (LAGBs) were defined as boundaries with a misorientation angle between 2° and 10°, while high-angle grain boundaries (HAGBs) were those with a misorientation angle exceeding 15°. Among the HAGBs, special attention was given to the twin boundaries (TBs), which were identified as boundaries exhibiting a misorientation of 60° ± 5° along the <111> crystallographic direction, represented with black lines in all orientation maps. This classification enables a detailed analysis of the role of different boundary types in the deformation mechanisms of the material, particularly in relation to grain rotation, strain accommodation, and twin formation during tensile testing.

Over time, high-temperature deformation leads to recovery mechanisms such as dislocation annihilation, subgrain formation, and recrystallization, as shown by the histograms of the grain size in Fig 7. Dislocation annihilation occurs when oppositely directed dislocations meet and cancel each other out, reducing the dislocation density. This process, along with subgrain formation, can contribute to the reduction in internal stresses and the creation of smaller, more equiaxed regions within the material. As recovery progresses, recrystallization may initiate, involving the formation of new, dislocation-free grains that supplant the deformed structure [61]. This process generally leads to a reduction in average grain size due to the finer nature of the recrystallized grains and have a direct impact on the grain size distribution, which is reflected in the EBSD histograms at different temperatures. During a high-temperature tensile test, grain size initially increases as a result of grain growth due to thermal activation, but as recrystallization progresses, the formation of new, smaller grains may counteract this grain coarsening effect. Consequently, the EBSD histogram shows smaller grain sizes as the temperature-induced recovery mechanisms evolve. The balance between grain growth and recrystallization will determine the final grain size distribution, which in turn affects the material's mechanical properties such as strength and ductility.

In order to provide a quantitative analysis of the EBSD images, we calculate the twin ratio as: Twin Ratio = $L_{TB}$/ Total $L_{GB}$ where $L_{TB}$ is the length of the TB and $L_{GB}$ is the total Length of the GBs. This ratio is shown in Fig 8 providing a direct measure of the extent to which twinning contributes to the overall GB network as a function of the temperature. The error bars in the plot represent the standard deviations of the twin ratios (as defined above), calculated from three orientation maps per test condition. A higher twin ratio suggests that twinning plays a significant role in strain hardening, with twin boundaries acting as obstacles to dislocation motion and contributing to the material's strengthening [25,26]. Conversely, a lower twin ratio indicates that the deformation is primarily dominated by dislocation slip and other mechanisms, such as dynamic recovery or recrystallization, which do not involve significant twin formation.

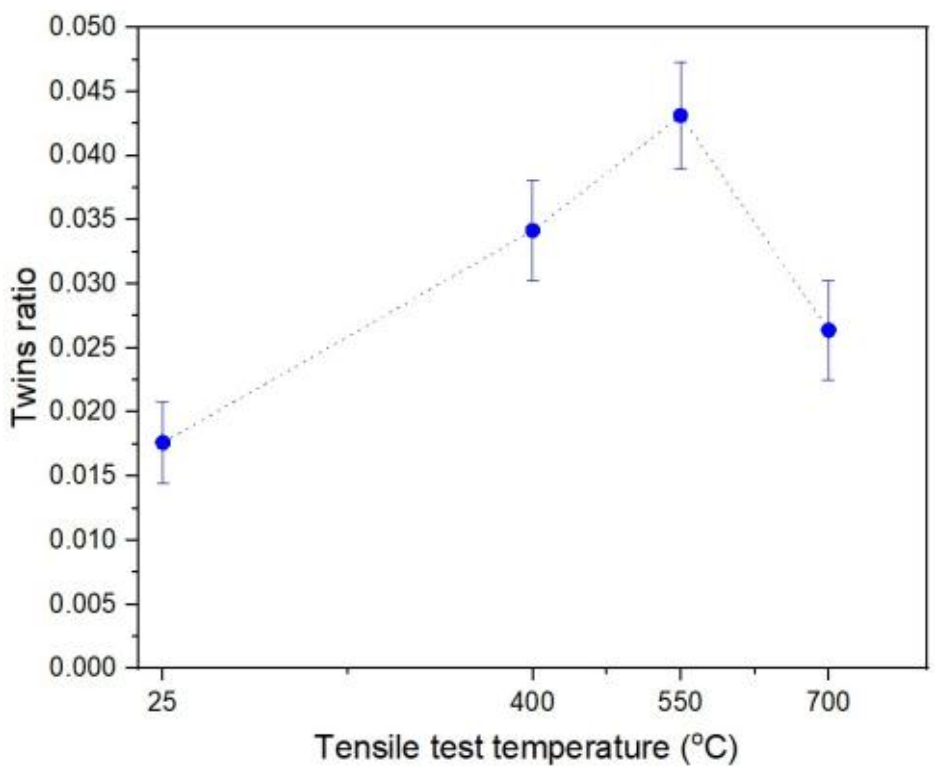

Figure 8. Fraction of twin boundaries in microstructure of HEA-1 alloy as a function tensile test temperature.

### 3.4. Computational modeling of twin formation mechanisms

In fcc HEA, plastic deformation is primarily accommodated by dislocation slip along the {111} planes in ⟨110⟩ directions, similar to conventional fcc metals [63]. The propensity for slip activation under an applied load is characterized by the Schmid factor, which quantifies the resolved shear stress acting on a given slip system. It is defined as: $m=\cos(\phi)\cdot\cos(\lambda)$, where $\phi$ is the angle between the loading direction and the normal to the slip plane, and $\lambda$ is the angle between the loading direction and the slip direction. During uniaxial tensile testing, slip is expected to initiate on the systems with the highest Schmid factor, making it a critical parameter for interpreting anisotropic deformation behavior. Although HEAs exhibit lattice distortions that may affect the critical resolved shear stress, the geometric relationship described by the Schmid factor remains valid. As such, it serves as a useful tool in computational studies to assess grain-scale mechanical response and the activation of specific slip systems under applied stress. Figure 9 presents atomic-scale maps of the Schmid factor at different strain levels and temperatures, providing a geometric measure of the propensity for slip system activation under applied load. These maps serve as a valuable tool for visualizing how microstructural features influence local deformation behavior. The orientation of slip planes relative to GBs and TBs critically governs dislocation motion and accumulation. High Schmid factor regions adjacent to GBs facilitate slip transmission across interfaces when the crystallographic alignment between neighboring grains is favorable. Conversely, GBs with large misorientations or poor alignment of active slip systems often act as strong barriers to dislocation glide, leading to local stress concentration and the potential nucleation of secondary deformation mechanisms such as grain boundary sliding or crack initiation. To complement these qualitative observations, the fraction of high Schmid factor regions ($Ds$), defined as the proportion of atoms with m>0.46 was computed as a function of temperature and strain. The results reveal a gradual increase in $Ds$ with both deformation and temperature, indicating enhanced slip compatibility and activation of favorably oriented regions at elevated temperatures. This quantitative trend supports the interpretation that thermally assisted local reorientation promotes easier slip transfer and strain accommodation across boundaries, consistent with the experimental evidence of increased ductility at high temperatures.

Twin boundaries, commonly formed during deformation in low-stacking-fault-energy fcc HEAs, interact with the active slip systems in a manner that depends on their crystallographic alignment as quantified by the Schmid factor. When slip systems are favorably oriented across twin boundaries, dislocations can transmit through the interface, although their mobility may be reduced due to the high coherency of the twin plane. Conversely, when the active slip systems terminate at the twin boundary or intersect it at unfavorable orientations, dislocation pile-ups occur, promoting local stress accumulation, work hardening, and, in some cases, the activation of secondary slip or twinning in adjacent grains. It should be emphasized that the Schmid factor maps represent geometric descriptors of the local crystallographic configuration relative to the applied load. Therefore, they do not directly reflect the temperature dependence of deformation mechanisms but rather indicate how changes in orientation and microstructural geometry under thermal and mechanical loading conditions influence the likelihood of slip activation. When analyzed together with the orientation and distribution of grain and twin boundaries, these maps provide a physically grounded framework for interpreting local slip activity, strain localization, and the overall mechanical response of fcc HEAs under applied stress.

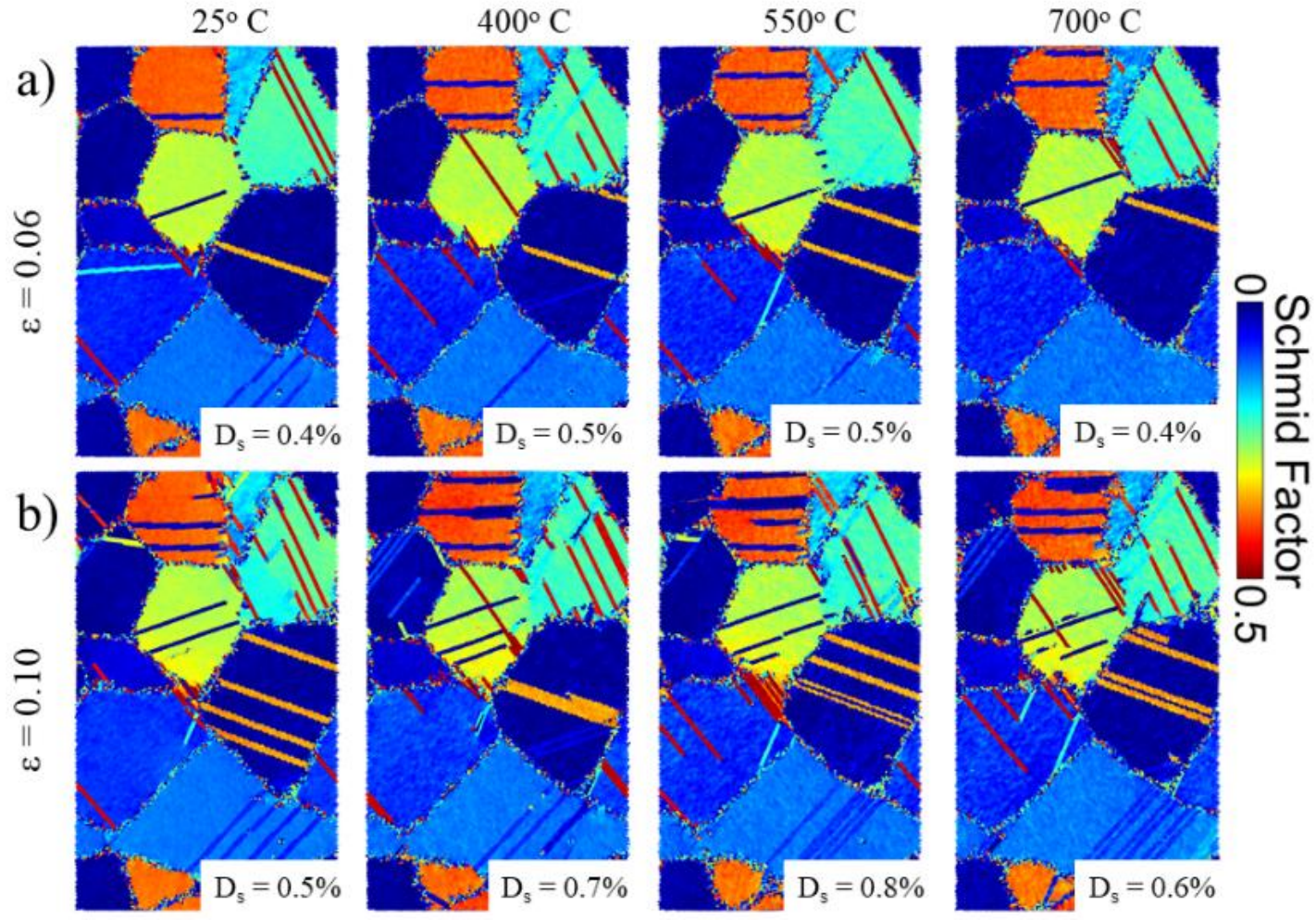

Fig 9. Atomic-scale maps of the Schmid factor at various strain levels and temperatures. The maps provide a geometric assessment of the propensity for slip system activation, highlighting how local orientation and loading conditions influence deformation behavior. The accompanying quantitative analysis ($D_s$) represents the fraction of high Schmid factor regions (m > 0.46), showing its evolution with temperature and strain.

In order to follow experimental conditions as close as possible to experimental measurements, MD simulations were conducted to model the cooling of HEA-1 for all the temperatures to room temperature with periodic boundary conditions in all directions and a timestep of 1 fs. The initial configuration consisted of a deformed microstructure containing dislocations, stacking faults, and random elemental distribution imported from a prior tensile test snapshot at 0.06 and 0.1 strain. Temperature and pressure control during cooling were achieved via a Nosé–Hoover thermostat and barostat, with damping parameters set to 0.1 ps and 1 ps, respectively. The system temperature was ramped linearly from a given temperature to room temperature over 100 ps to prevent thermal shock and capture thermal compression accurately under constant pressure. Once room temperature was reached, the sample was held at this temperature under NPT conditions for 2 ns to permit thermally activated processes [23-26]. Thermodynamic data (temperature, pressure, total energy) were logged every 1 ps, and atomic configurations were saved at intervals of 50 ps for structural analysis. In Fig 10a), we present results for the number of HCP atoms as a function of the applied strain at different temperatures. Under applied strain, perfect fcc regions undergo dislocation nucleation and motion, where full dislocations frequently dissociate into Shockley partial dislocations, leading to the formation of intrinsic and extrinsic stacking faults [23]. These stacking faults locally transform the fcc structure into an HCP configuration, and their accumulation may promote deformation twinning, further contributing to the growth of HCP regions, as observed in Fig 9. In highly deformed zones, particularly at high strain rates, local atomic rearrangements can result in the transient formation of BCC-like structures, indicative of severe lattice distortions or phase transitions under extreme stress states. Additionally, dynamic recovery and recrystallization processes may partially restore the fcc structure as dislocations annihilate or rearrange into low-energy configurations.

During tensile testing of fcc metals, twin boundaries can form as a result of plastic deformation, particularly when dislocation motion is hindered by obstacles such as grain boundaries. The stress-strain response of the HEA-1 shown in Fig 5 at 550°C exhibits noticeable stress drops at various strain levels, which can be attributed to the interplay between deformation twinning and dynamic recrystallization (DRX), this process is thermally activated, leading to the nucleation and growth of strain-free grains, primarily at grain boundaries and highly deformed regions. The onset of DRX results in a reduction in the overall dislocation density, temporarily decreasing the flow stress and producing characteristic stress drops. The cyclic nature of these stress fluctuations suggests a dynamic balance between strain accumulation from dislocation activity and twinning, and strain relaxation through recrystallization events. In order to understand this mechanism from MD simulations, in Fig 10 we present the visualization of the formation of planar faults with dislocation in b) and fcc atoms with planar faults in c) in the HEA-1 at a temperature of 550ºC at a strain of 0.06. The classification of planar faults is carried out with Ovito, where the identified HCP atoms that are arranged on parallel {111} planes of the fcc crystal are classified in intrinsic planar stacking fault with an order as ABACAC or two adjacent hcp-like layers, the coherent twin boundary ABCACB or one hcp-like layer, and the multilayer stacking fault can be a combination of the previous ones or three or more adjacent hcp-like layers. As tensile stress is applied, dislocations move through the crystal lattice, but their motion may become obstructed by grain boundaries, leading to localized strain accumulation, as shown in Fig 10b). In such regions, where dislocations pile up, twinning can occur as an alternative deformation mechanism. The formation of twin boundaries causes a mirror-image displacement of atomic planes across the boundary as observed in Fig 10c). Grain boundaries themselves can act as nucleation sites for twin dislocations, and their presence increases the likelihood of twinning, especially under high strain conditions. Additionally, at fine grains, the higher density of grain boundaries facilitates twin formation due to increased stress concentration. The temperature, strain rate, and grain size are key factors influencing the extent of twinning, at higher temperatures or lower strain rates, dislocation motion dominates, and twinning may be less significant. Therefore, in experiments, several mechanisms may be overlooked due to the limitations of microscopy observations or the post-mortem analysis of samples, whereas MD simulations enable the investigation of slip dislocation, twin formation, and grain dynamics under conditions that closely approximate experimental environments.

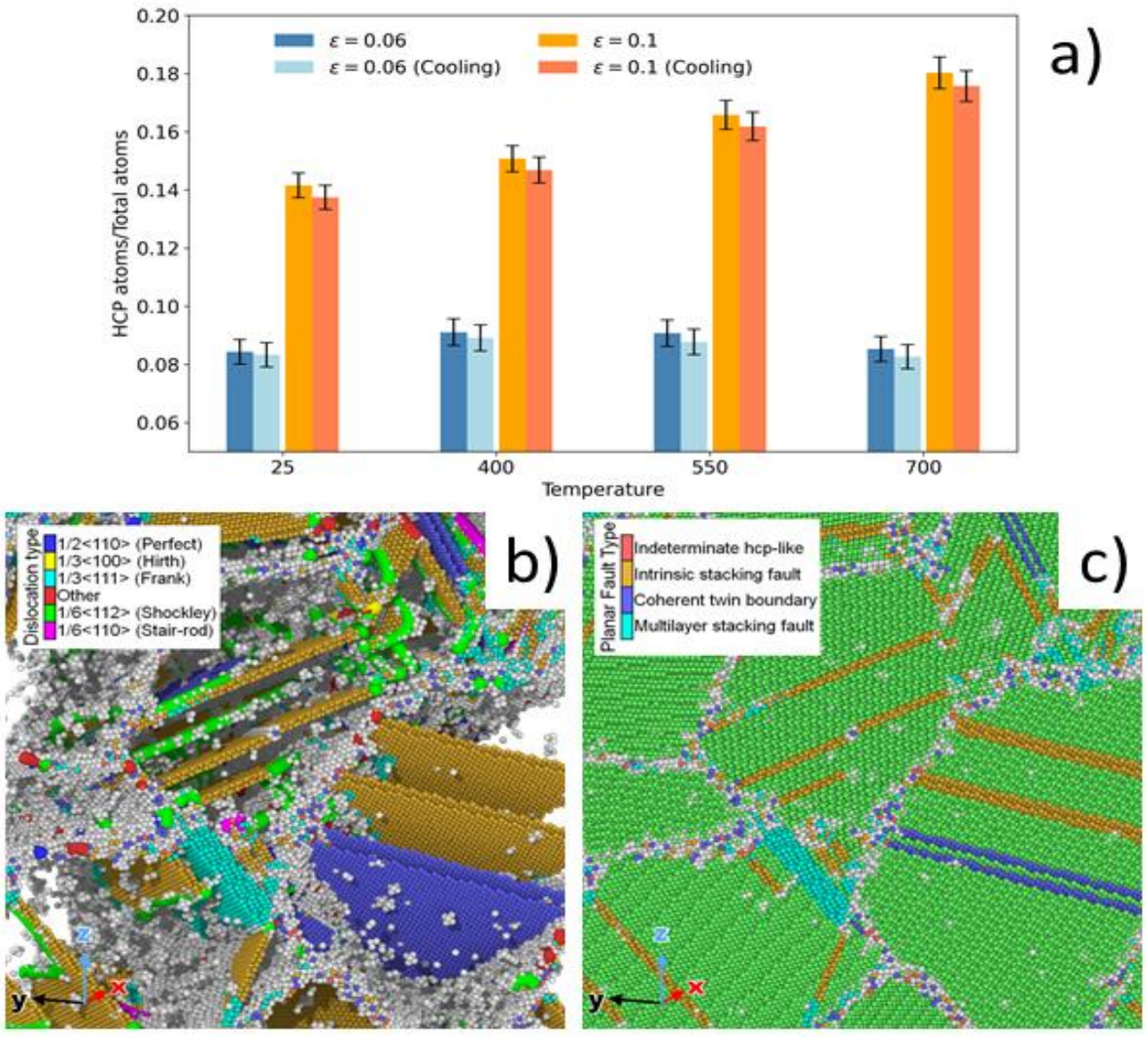

Fig 10. Visualization of planar fault types after identifying HCP atoms in the HEA-1 at 550 °C and a strain of 0.06. (a) Evolution of the number of HCP and BCC atoms as a function of the applied strain at different temperatures. (b) Atomic-scale visualization of dislocation formation around intrinsic stacking faults, highlighting the volumetric characterization of the grain. (c) Planar fault structures showing the spatial distribution and morphology of stacking faults within the deformed microstructure. The fcc atoms in (b) and (c) are displayed as cross-sectional cuts along the x-axis to reveal the internal faulted regions.

Fig. 11 presents the identification of twin boundaries after the cooling process for each tensile test conducted at different temperatures, at strains of 0.06 (a) and 0.1 (b), selected to best match the conditions of the EBSD images. Faulted planes were identified using OVITO, and intrinsic stacking faults were excluded from the MD visualizations to isolate the twin boundaries. Thus, the visibility and apparent number of twin boundaries depend strongly on the orientation of the sample cross-section and the specific strain at which the MD snapshot is taken. Consequently, the quantification of twin boundaries may vary depending on the viewing plane and the selected deformation stage [62].

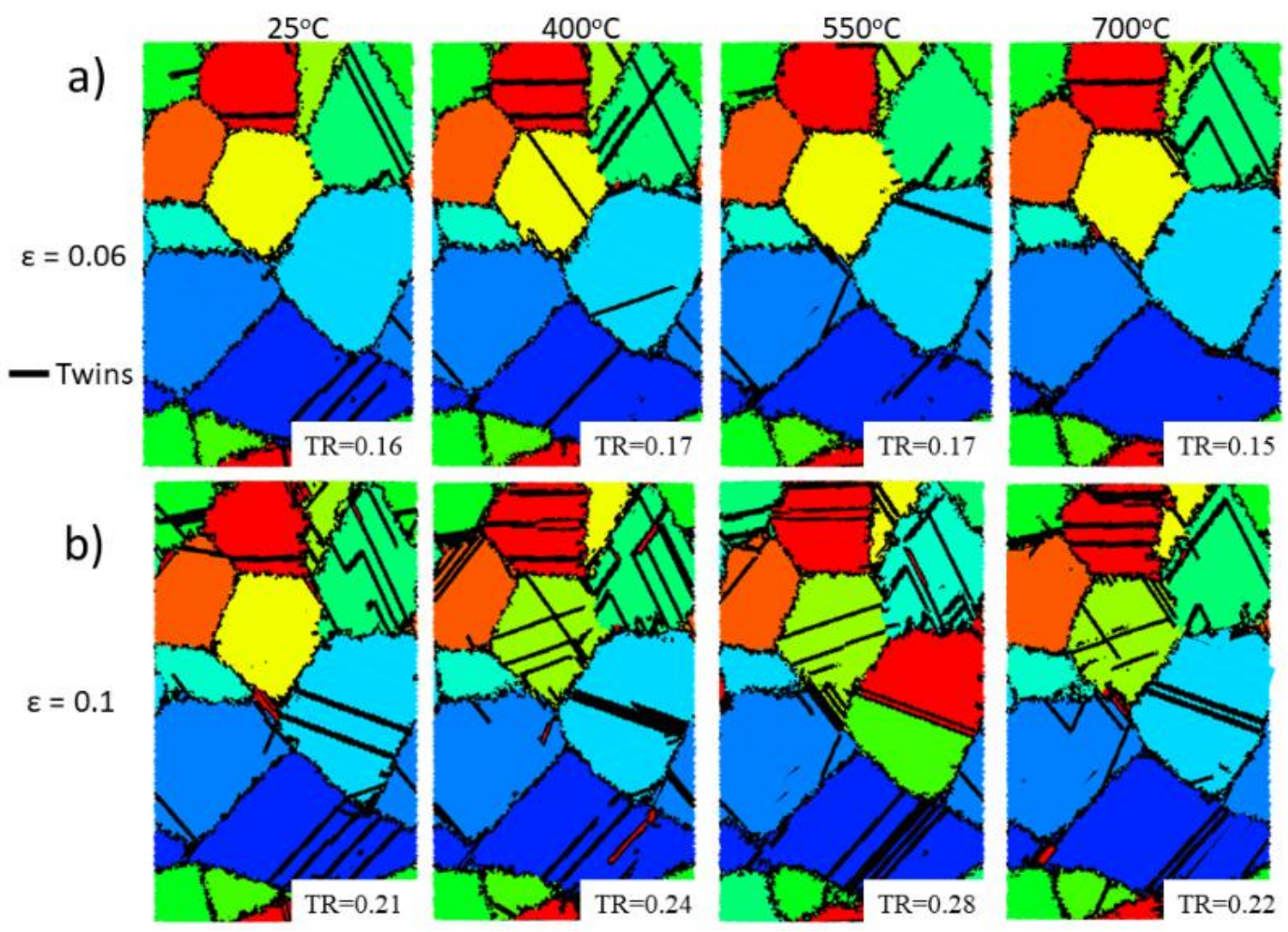

Fig. 11 MD simulations at strains of ε = 0.06 (a) and ε = 0.10 (b) for different temperatures, showing the nucleation and evolution of stacking-fault planes and twin boundaries within the grains. The corresponding twin ratio (TR), defined as the ratio of the total twin plane area to the total grain boundary area, is also shown to quantify the extent of twinning at each strain and temperature.

By removing the intrinsic stacking-fault planes from the MD results, we quantified the formation of twin boundaries at different strain levels. To enable a consistent comparison with the experimental measurements, we computed the twin ratio (TR) as the ratio of the total area of identified twin planes to the total grain boundary area. At a strain of 0.06, corresponding to the onset of plasticity for all temperatures, the TR value is approximately 0.16 at room temperature and presents minimal changes across all temperatures. At a higher strain of 0.10, the TR increases to 0.21 at room temperature, 0.24 at 400 °C, 0.28 at 550 °C, and 0.22 at 700 °C, indicating enhanced twin boundary formation at intermediate temperatures. This trend follows the experimental observations shown in Figure 7, where a higher density of twins was also observed at 400 °C and 550 °C, while both the room-temperature and high-temperature conditions exhibited lower values. These results confirm the consistency between the atomistic and experimental analyses and highlight the temperature-dependent nature of twin formation in the HEA-1 alloy.

It is worth noting that MD simulations inherently operate at nanometric length scales and high strain rates (on the order of $10^9$ $s^{-1}$), which impose physical constraints on the deformation mechanisms that can be realistically captured [64]. Under such conditions, thermally activated processes such as dislocation climb, dynamic recovery, and recrystallization are effectively suppressed due to the limited simulation timescales. Conversely, fast deformation pathways, including partial dislocation glide, stacking-fault formation, and deformation twinning, are preferentially activated, resulting in their apparent dominance in the atomistic simulations. Consequently, while the present MD results provide valuable insight into the fundamental atomic-scale processes occurring within the non-equiatomic NiFeCrMn alloy, the relative significance of each mechanism under experimental strain rates must be interpreted at some extent. In this context, the simulations serve primarily to complement the experimental tensile and EBSD analyses, by identifying the plausible defect interactions and local transformation modes that contribute to plasticity. The observation of twinning in MD simulations is therefore interpreted as an indicator of potential deformation pathways rather than as the exclusive mechanism governing macroscopic response. This integrative approach, combining high-resolution experimental characterization with atomistic modeling, enables a more comprehensive understanding of the deformation behavior of Co-free high-entropy alloys and provides a physically grounded framework for interpreting the temperature-dependent mechanical response observed experimentally.

## 4. Conclusions

Cobalt-free high-entropy alloys (HEAs) are emerging as strong candidates for nuclear structural applications due to their excellent mechanical performance, thermal stability, and resistance to radiation damage, while eliminating concerns associated with long-lived Co radioisotopes. In this study, we investigated the plastic deformation behavior of an fcc CrMnFeNi-based HEA ($Cr_{15.5}Mn_{13.7}Fe_{39.2}Ni_{31.6}$) through a combined experimental and computational approach under tensile loading at room and elevated temperatures. Tensile testing revealed a clear temperature-dependent reduction in mechanical strength, which we attributed to the activation of thermally driven deformation mechanisms and associated microstructural changes. Molecular dynamics simulations of both single- and polycrystalline configurations provided atomistic insights into stacking fault evolution and dynamic grain behavior. The formation and evolution of planar defects, including twin boundaries, were correlated with strain and temperature, offering a mechanistic link between atomic-scale processes and macroscopic mechanical response. EBSD confirmed twin formation and grain boundary activity at elevated temperatures, in agreement with the simulation results. Grain orientation played a critical role, with [111]-oriented grains showing a gradual decrease in strength and [110]-oriented grains exhibiting significant softening above 550 °C. Furthermore, the absence of cobalt was found to influence thermally activated deformation pathways, contributing to an increase in ultimate tensile strength at high temperatures compared to the canonical Cantor alloy. Moreover, we report that the higher Ni content in HEA-1 increases its SFE, enhancing dislocation pinning and yield strength. Analysis of local orientations using the Schmid factor, which quantifies the resolved shear stress acting on specific slip systems, further elucidated the anisotropic deformation response. Despite the inherent lattice distortions and chemical complexity of HEAs, the Schmid law remains applicable, providing a reliable geometric framework to interpret slip system activation and grain-dependent softening. These results emphasize the role of local SFE fluctuations and elemental tuning, especially of Ni and Co, in governing deformation mechanisms in concentrated alloys.

The observed decrease in ductility with increasing temperature can be attributed to the competing effects of thermally activated deformation and recovery mechanisms. At moderate temperatures (400–550 °C), the activation of partial dislocations and twin formation enhances strain accommodation through localized plasticity. However, at higher temperatures, the increased atomic mobility promotes dynamic recovery, leading to the annihilation and rearrangement of dislocations into low-energy configurations. This process suppresses dislocation storage and reduces the strain-hardening capability of the alloy. Additionally, grain boundary sliding and local stress relaxation become more pronounced at elevated temperatures, promoting early strain localization and intergranular damage. Consequently, while strength decreases smoothly with temperature due to thermal softening, the reduction in work-hardening rate and the onset of boundary-mediated deformation collectively result in a lower strain to fracture. This mechanistic interpretation aligns with both the experimental observations and the MD-predicted microstructural evolution across the studied temperature range. Overall, this work enhances the fundamental understanding of deformation mechanisms in Co-free HEAs and supports their continued development as radiation-resistant structural materials for advanced nuclear energy systems.

**Acknowledgments**

Research was funded through the INNUMAT (Grant Agreement No. 101061241) and European Union Horizon 2020 research and innovation program under Grant Agreement No. 857470 and from the European Regional Development Fund under the program of the Foundation for Polish Science International Research Agenda PLUS, Grant No. MAB PLUS/2018/8, and the initiative of the Ministry of Science and Higher Education "Support for the activities of Centers of Excellence established in Poland under the Horizon 2020 program" under Agreement No. MEiN/2023/DIR/3795. The Ministry of Science, Technological Development, and Innovation of the Republic of Serbia, grant No. 451-03-136/2025-03/200023. We gratefully acknowledge Polish high-performance computing infrastructure PLGrid (HPC Center: ACK Cyfronet AGH) for providing computer facilities and support within computational Grant No. PLG/2024/017084. T.K. acknowledges support received from Project No. 2022/47/P/ ST5/01169 co-funded by the National Science Centre and the European Union Framework Programme for Research and Innovation Horizon 2020 under the Marie Skłodowska-Curie Grant Agreement No. 945339. One of the authors (AD) acknowledged to have received support from the French ANR-PRCE-HERIA project (ANR-19-CE08-0012-01) to develop the EAM potential. The studied alloy was developped under co-funding by EDF company (2018, a master thesis work by S. Boulila), one of the authors (AF) acknowledged this financial support.